\documentclass{pCHEF}

\title{Arbor, a new approach of the Particle Flow Algorithm}

\ShortTitle{Arbor PFA}

\author{\speaker{Manqi Ruan}\thanks{}\\
        Laboratoire Leprince-Ringuet, \'Ecole polytechnique, CNRS/IN2P3, Palaiseau, France\\
	IHEP, Beijing, China
        E-mail: \email{Manqi.ruan@ihep.ac.cn}}
\author{Henri Videau\\
        Laboratoire Leprince-Ringuet, \'Ecole polytechnique, CNRS/IN2P3, Palaiseau, France\\
        E-mail: \email{videau@llr.in2p3.fr}}

\abstract{

The granularity of calorimeter has been revolutionary boosted for future collider experiments. 
The calorimeter has been pushed to a stage that the sub structure of showers especially hadronic showers can be recorded to a high precision. 
New reconstruction algorithms are expected from these informations. 

Following the idea that shower follows the topology of the tree, we developed Arbor, a Particle Flow Algorithm framework. 
Tested on both simulated data and test beam data, it can successfully separate nearby showers. 
It has comparable jet energy resolution the best PFA algorithm for International Linear Collider. 
More importantly, Arbor successfully tags the sub shower structure such as the trajectory of charged particles generated in shower cascade, 
enabling new approaches for event reconstruction with high granularity calorimeter.
}

\FullConference{Calorimetry for High Energy Frontiers - CHEF 2013,\\
		April 22-25, 2013\\
		Paris, France}

\usepackage{CHEF2013_definitions}
\begin{document}

\section{Introduction}

The discovery of Higgs boson at LHC brings the experimental high energy physics into a new era.
The central tasks of the coming decades are the precise measurement of Higgs properties and hunting for new physics signals. 
The jet energy resolution is crucial for both tasks. 

The concept of Particle Flow Algorithm has been introduced to achieve excellent jet energy resolution. 
The idea is based on the fact that most of the jet energy distributes in charged particles, whose momentum can be measured to a precision much better than the calorimeter. 
Therefore, by separating and reconstruct each jet particle in the most suited sub-detectors (charged particle at tracker, photons at ECAL and neutral hadrons at ECAL and HCAL),
the jet energy can be measured to a precision much better than the conventional method, which is to measure the jet energy using only calorimeters. 
It has already achieved great success in both ALEPH and CMS experiments, where the jet energy resolution can be measured to roughly a precision of $50-60\%/\sqrt{E_{J}}$.

The $e^{+}e^{-}$ machines such as ILC, CLIC, TLEP or CEPC are regarded as ideal machine for precise measurement of Higgs properties, while it also has very distinguishable advantages in new physics hunting comparing to the hadron machines, namely free of QCD background and has precisely measureable and adjustable initial states.
The objective jet energy resolution at the $e^{+}e^{-}$ machines is usually referred to as $3\%$ of relative accuracy, roughly a factor of two better than what had been achieved at ALEPH and CMS.
Such accuracy is needed to distinguish the Z boson from W boson in their hadronic decay mode, while it also provides a much better separation between the Z boson and Higgs boson. 

\begin{figure}
\begin{center}
\includegraphics[width=.8\textwidth]{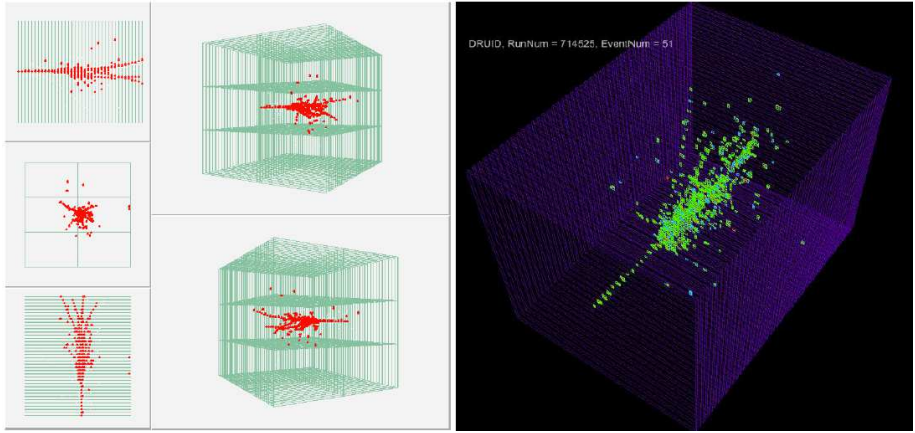}
\caption{Hadronic shower at CALICE GRPC DHCAL~\cite{DHCAL} and SDHCAL~\cite{SDHCAL} Prototypes}
\label{EvtDisp}
\end{center}
\end{figure}

On the other hand, the development of micro electronic technology can highly boost the granularity of the detector readout up to 3 orders of magnitude (Fig.~\ref{EvtDisp}).
The total number of electronic channels at SiD/ILD calorimeter reaches $10^{8}$, or equivalently a density of $1/cm^{3}$.
Such high granularity is favored by the particle flow algorithm where the separation between different jet particles is crucial,  
while the event shape and more importantly the shower spatial development are recorded to an unprecedented precision.
In fact, such calorimeter brings us back the detail of shower forgotten since the age of bubble chamber, and therefore also been called 3-d or even 5-d (with time and energy deposition information) imaging calorimeter. 

\begin{figure}
\begin{center}
\includegraphics[width=.8\textwidth]{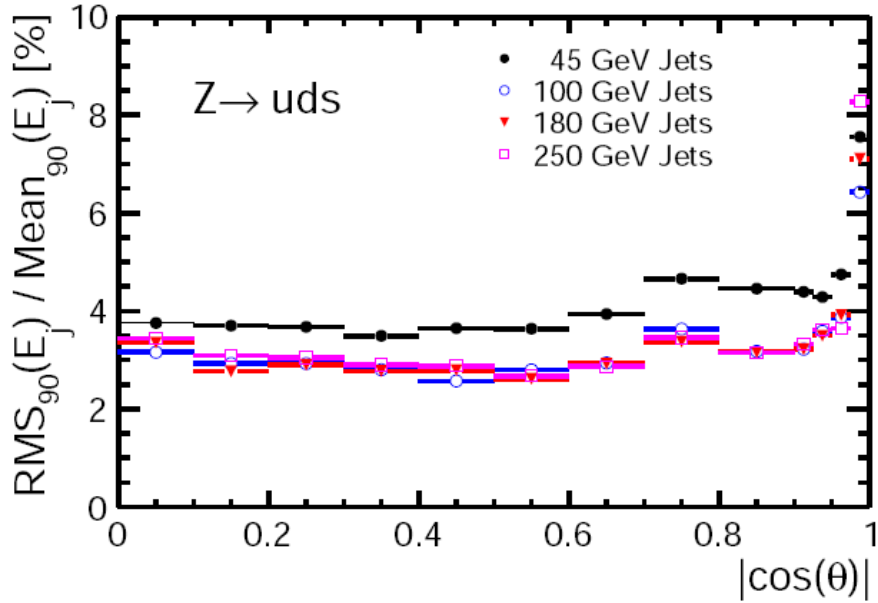}
\caption{Jet energy resolution achieved by PandoraPFA}
\label{PandoraPerformance}
\end{center}
\end{figure}

To improve the jet energy resolution by a factor of two is not an easy task. 
However, since the $e^{+}e^{-}$ machine is free of QCD background and pileup comparing to the LHC, 
while the imaging calorimeter records much more information, 
this goal is still within the reach with advanced algorithm development. 
People tried many different algorithms, while currently the best algorithm, PandoraPFA~\cite{PFApaper}, has already achieved such precision (Fig.~\ref{PandoraPerformance}). 

In this paper, we would like to present another PFA algorithm, namely Arbor algorithm. 
The idea is not only to re-validate the principle of PFA, 
but also to fully investigate the reconstruction potential from ultra-high granularity calorimeter. 

\section{Arbor Algorithms: basic principle}

\begin{figure}
\begin{center}
\includegraphics[width=.8\textwidth]{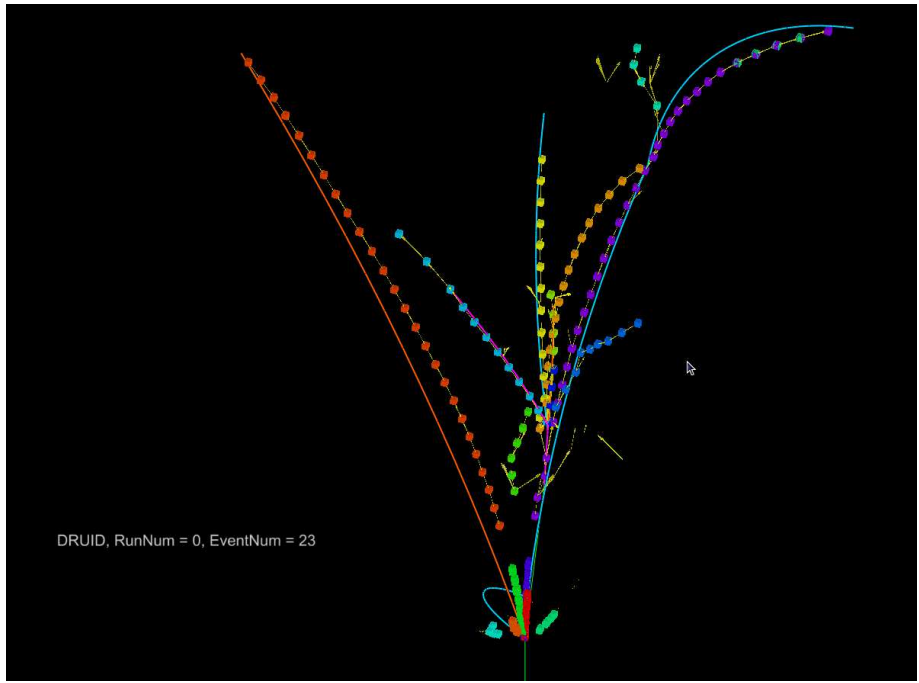}
\caption{A 20~GeV $K^{0}_{L}$ shower reconstructed by Arbor Algorithm}
\label{ArborPrinciple}
\end{center}
\end{figure}

Arbor algorithm, as its name suggests, is inspired by the fact that the shower spatial development follows the topology of a tree. 
At an imaging calorimeter, the spatial configuration of particle showers especially hadronic showers are recorded to such a precision that we could use Arbor algorithm to reconstruct its tree structure, see Fig.~\ref{ArborPrinciple}.
Arbor reconstruct the long travelling charged particles generated at hadronic shower as the branches of the tree, while the tree structure is ensured with the constrain that no loop structure is allowed. 

In the following context, we will present in details that how this algorithm is realized. 

\section{Core Algorithm}

In its current configuration, Arbor use only geometry information of the hits, i.e, no energy information is used. 
After necessary hit cleaning, 
if the distance between any pair of hits is smaller than a given threshold, a local connectors is build. 
The connector is an orientated arrow which links a pair of hits and end at the hit with larger transverse distance to the origin. 

\begin{figure}
\begin{center}
\includegraphics[width=.8\textwidth]{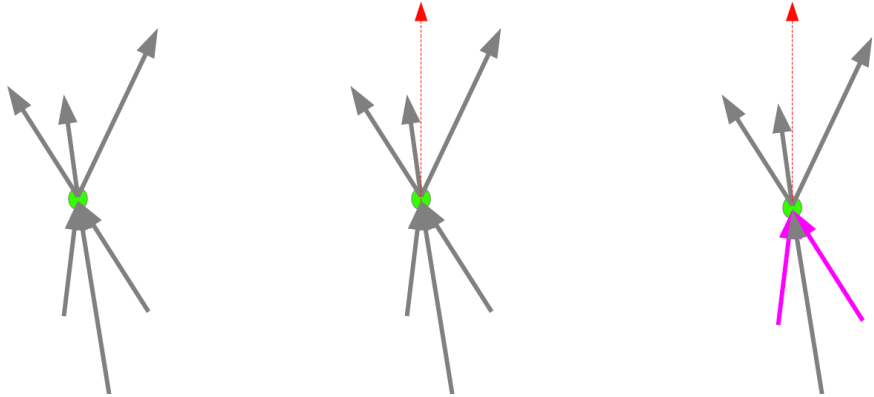}
\caption{Arbor cleaning procedure of the connector. the left plot demonstrate a hit with all its connectors; the red arrow in the middle plot indicate the reference direction, while the pink arrows in the right plot indicate the connectors that will be dropped by the cleaning procedure}
\label{Conn}
\end{center}
\end{figure}

The second step is to clean the connectors, see Fig.~\ref{Conn}.
After the first step, there can be multiple connectors end or begin at a given hit.
Using the directions and length of these connectors as well as the spatial position of the hit, 
a reference direction can be calculated. 
For all the connectors end at this hit, Arobr kept at most one connector that has the minimal angle to the reference direction. 
Therefore, no loop structure will be kept after the cleaning, and a tree structure based on the connectors emerges.

The connector's tree structure can be iterated: new connectors can be added according to the relative positions between hits as well as their reference directions, and the set of connectors can always been cleaned with similar criteria. 
The propose of the iteration is simply to find the best connector configurations, in the sense that every branch should be as smooth as possible and allowance of long connectors. 

\begin{figure}
\begin{center}
\includegraphics[width=.8\textwidth]{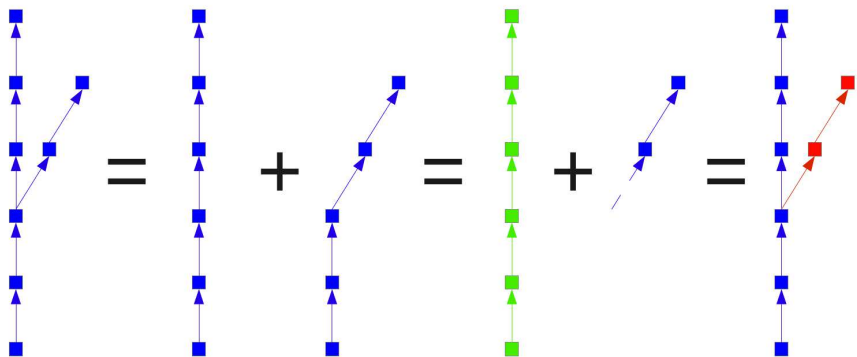}
\caption{Decouple tree structures into branches}
\label{Branch}
\end{center}
\end{figure}

Given a connector configuration follows the tree topology, we can define a hit as leaf is there are connectors end at it but no connector begin at it. 
To the contrary, a hit is called a seed if there are only connector(s) begin(s) at it. 
From each leaves, there is an unique path to trace back to one seed.
Sort these paths by their length, start from the longest path and flag this path as a branch, then trace along the second path until it ends up at the seed or at the hits belongs to the first branch.  
This procedure, demonstrated in Fig.~\ref{Branch}, decouples the tree structure into sets of branches, which represent the trajectories of charged particles.
Each branches has two reference directions, namely the one at its seed and at its leaf.
Those directions provides extra information for us to link branches or say tree into the shower clusters, corresponding to individual incident particles.  

\section{Performance}

In this section, we will introduce the performance of Arbor at different levels. 

\subsection{Sub shower level performance}

\begin{figure}
\begin{center}
\includegraphics[width=.8\textwidth]{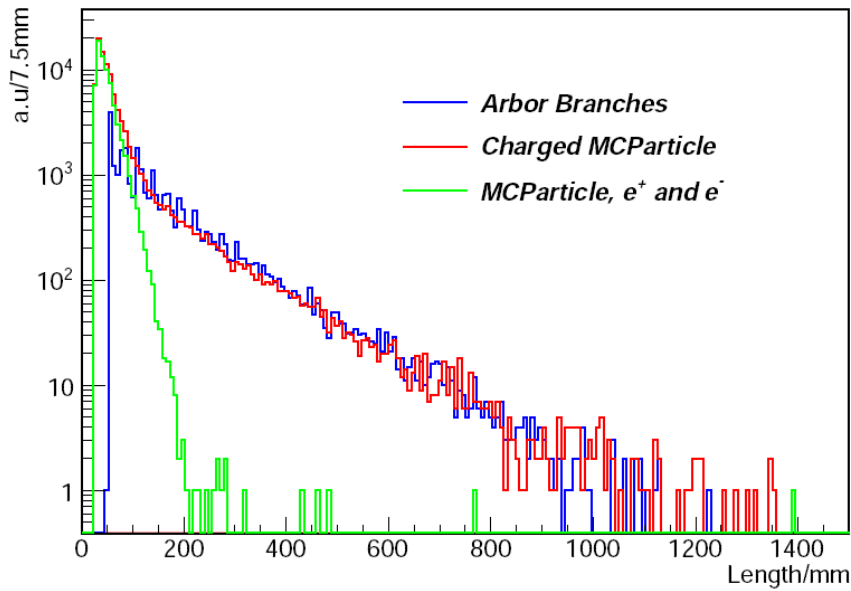}
\caption{Branch Length Vs Charged Particle Trajectory Length}
\label{Length}
\end{center}
\end{figure}

Introduced in above section, Arbor reconstructs the trajectories of charged particles into tree branches.
Therefore, we can compare the length of these two at Monte-Carlo simulation, see Fig.~\ref{Length}.   

The red curve in Fig.~\ref{Length} represent the length of MC particle trajectories, in terms of the spatial distance between generating point and end point. 
It can be decoupled into two populations, the electromagnetic trajectories deposited by $e^{+}$ or $e^{-}$ (green curve) and the hadronic trajectories. 
The blue curve shows the length of Arbor reconstructed branches, in terms of the sum of distance between neighboring hits. 
The red and blue curves agrees nicely for the length larger than 100~mm. 
In other word, Arbor can successfully tag the sub-shower trajectories.

\subsection{Shower level performance}

\begin{figure}
\begin{center}
\includegraphics[width=.95\textwidth]{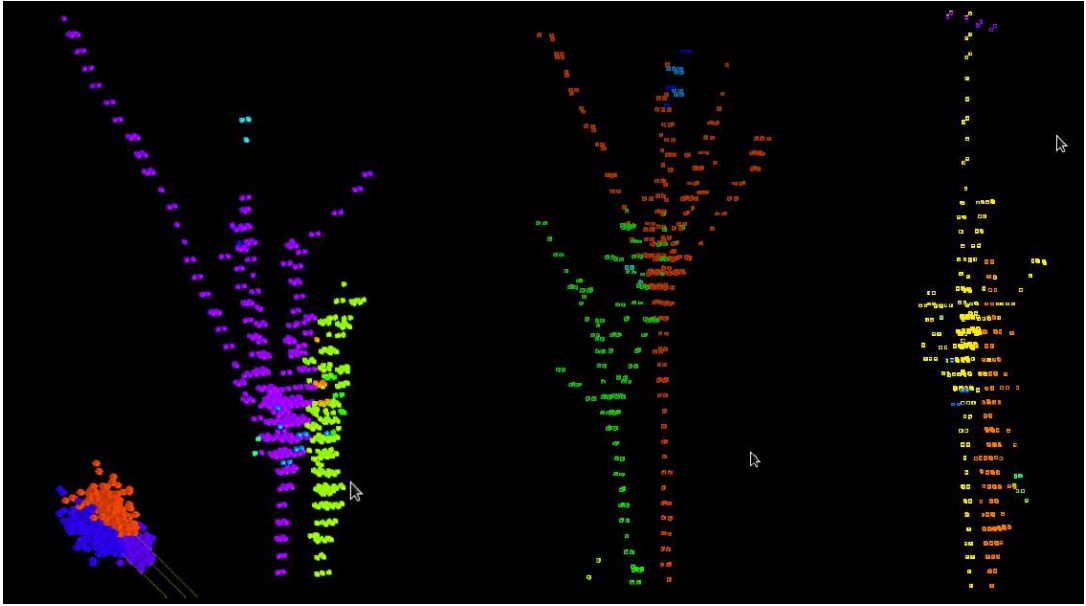}
\caption{Nearby showers reconstructed by Arbor. The display at left corner shows three nearby photon clusters, while the other three display shows nearby hadron showers }
\label{NbShower}
\end{center}
\end{figure}

In the ideal case, each tree represents the shower of one particle. 
Since the seed and the tree has one one correspondence, Arbor is a powerful algorithm to separate nearby showers, which is highly appreciated at Particle Flow Algorithm. 
Fig.~\ref{NbShower} shows some examples, where nearby showers are clearly separated by Arbor.

\begin{figure}
\begin{center}
\includegraphics[width=.6\textwidth]{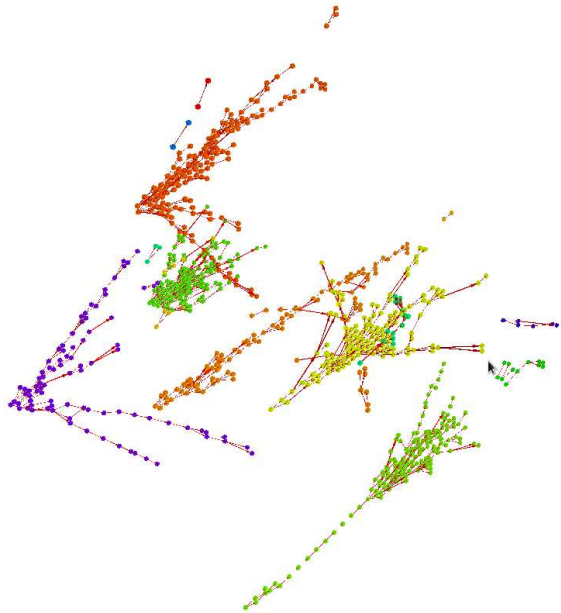}
\caption{A pre-interaction hadronic event recorded at test beam data, reconstructed with Arbor}
\label{CaliceShower}
\end{center}
\end{figure}

Arobr algorithm has also been applied to test beam data, where close showers are also easily separated, see Fig.~\ref{CaliceShower}.

\subsection{Jet level}

\begin{figure}
\begin{center}
\includegraphics[width=.98\textwidth]{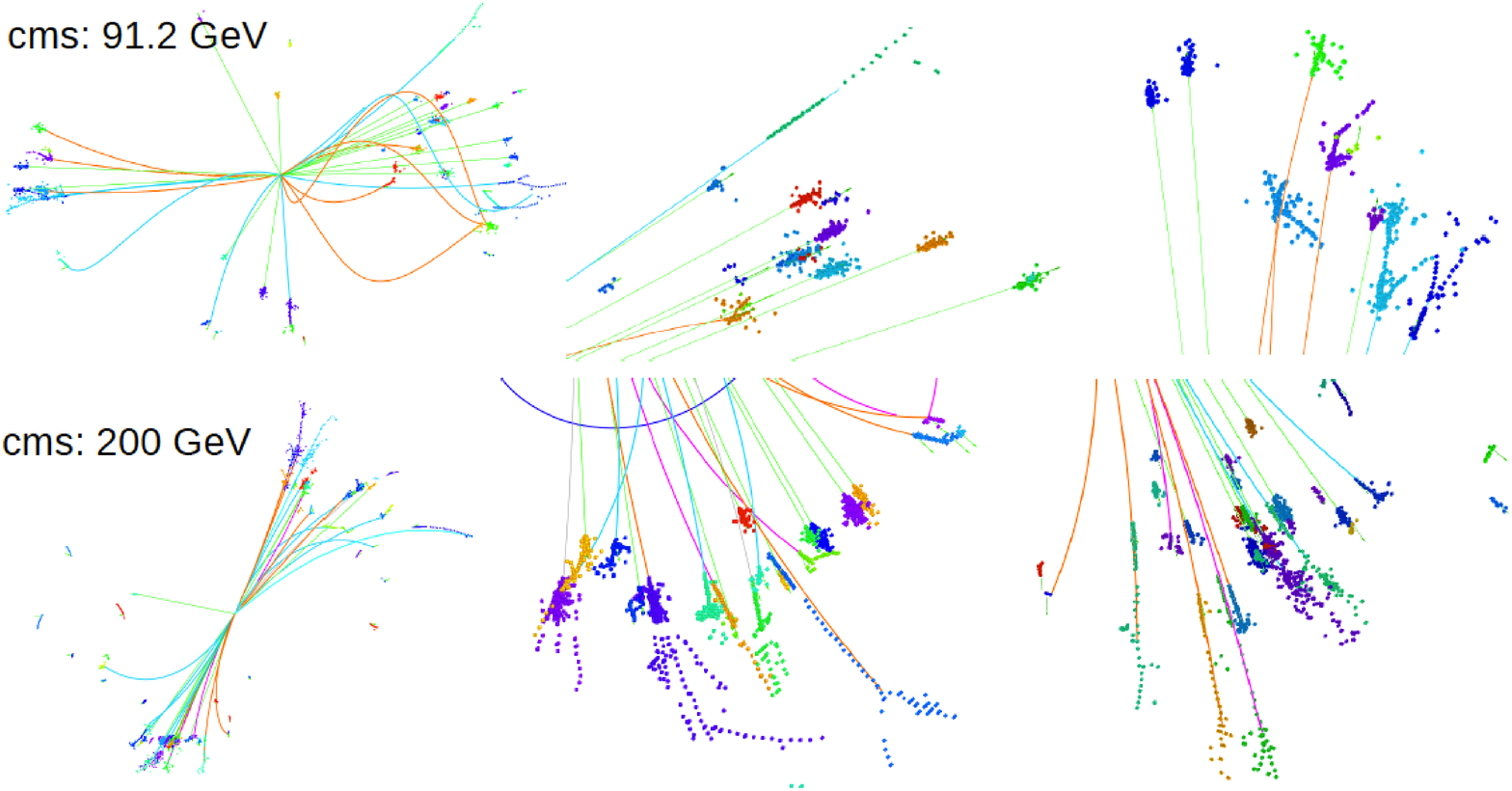}
\caption{QQ events reconstructed with Arbor. Above plots corresponding to qq event at Z threshold, below shows that at center of mass energy of 200~GeV}
\label{ArborQQ}
\end{center}
\end{figure}

Since Arbor can successfully separate nearby showers, it can be used to reconstruct the jet energy, see Fig.~\ref{ArborQQ}.

\begin{figure}
\begin{center}
\includegraphics[width=.98\textwidth]{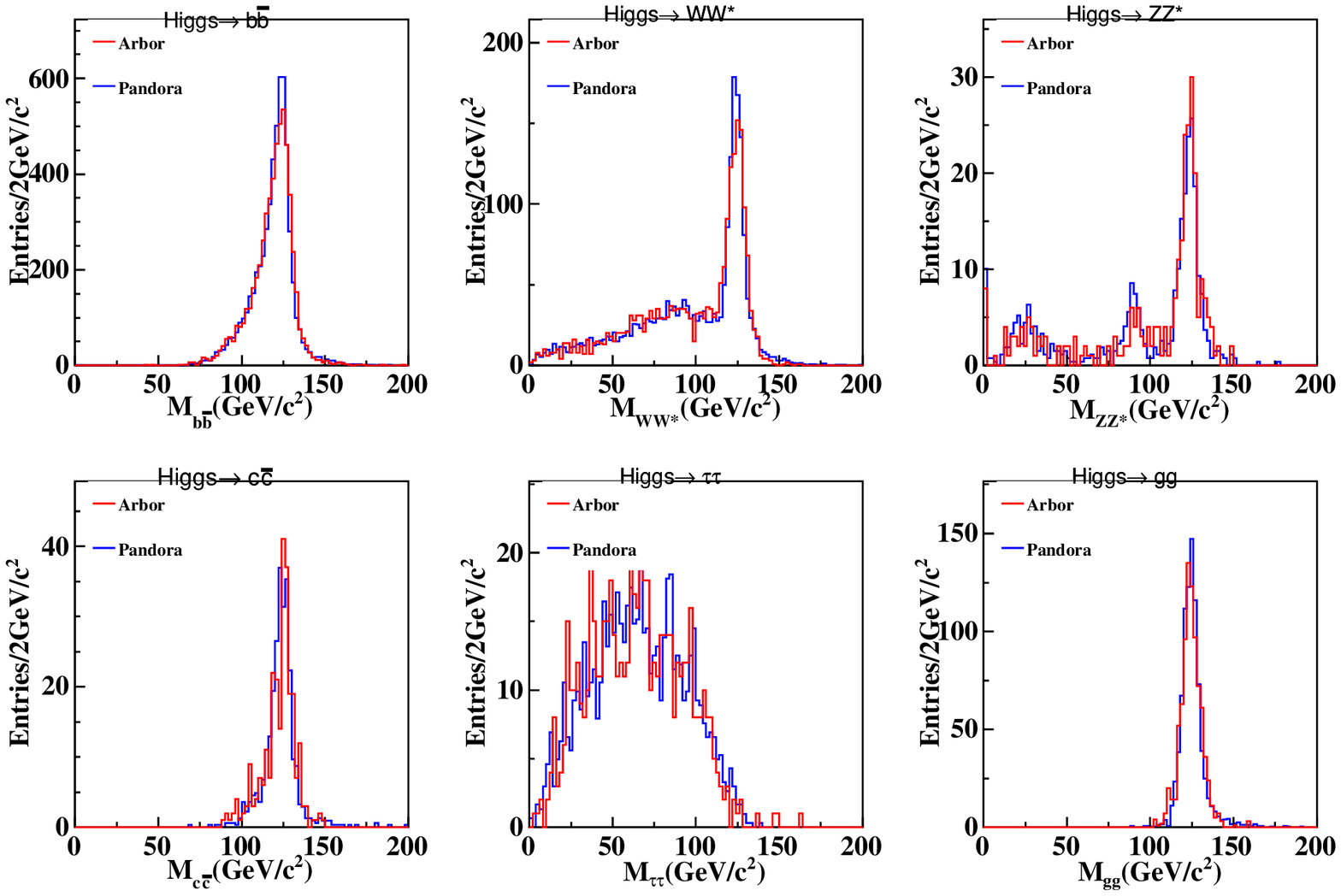}
\caption{Total reconstructed energy of vvH events, with Higgs decay into different final states. }
\label{Perform}
\end{center}
\end{figure}

Fig.~\ref{Perform} shows the total reconstructed energy for vvH events with both Arbor (red curve) and PandoraPFA (blue curve) with different Higgs decay final states.
There are marginal difference between the simulation sample: 
for Arbor, the samples are simulated with ILD detector equipped with DHCAL, where the HCAL works at a digital mode. 
for Pandora, the samples are simulated with AHCAL-equipped ILD detector, whose intrinsic hadronic shower energy resolution is better than DHCAL worked with hit counting method. 
Therefore, in terms of jet energy resolution, Arbor is comparable with PandoraPFA.

\section{Summary}

Following the idea that showers follows the topology of a tree, 
a new particle flow algorithm toward the high granularity calorimeter system, Arbor, has been developed. 
It has a strong separation power on nearby showers, which is crucial for the jet energy resolution.
Tested on qq event, the jet energy resolution reaches a level that can be compared to PandoraPFA. 
More imporantly, it can successfully tag the sub shower structures, for example, the trajectories of charged particles. 
Such capability provides huge potential for the reconstruction performance. 


\bibliography{mybib}{}
\bibliographystyle{unsrt}

\end{document}